\documentclass{vgtc}                   

\ifpdf
  \pdfoutput=1\relax                   
  \usepackage{graphicx}                
  \DeclareGraphicsExtensions{.pdf,.png,.jpg,.jpeg} 
\else
  \usepackage{graphicx}                
  \DeclareGraphicsExtensions{.eps}     
\fi%

\graphicspath{{figures/}{pictures/}{images/}{./}} 

\usepackage{microtype}                 
\PassOptionsToPackage{warn}{textcomp}  
\usepackage{textcomp}                  
\usepackage{mathptmx}                  
\usepackage{times}                     
\usepackage{cite}                      
\usepackage{tabu}                      
\usepackage{booktabs}                  
\onlineid{0}
\vgtccategory{Research}
\vgtcinsertpkg

\title{Visualizing a Large Spatiotemporal
Collection of Historic Photography with a Generous Interface}

\author{Taylor Arnold\thanks{e-mail: tarnold2@richmond.edu}\\ %
     \scriptsize University of Richmond, Math and Computer Science %
\and Nathaniel Ayers\thanks{e-mail: nayers@richmond.edu}\\ %
     \scriptsize University of Richmond, Digital Scholarship Lab %
\and Justin Madron\thanks{e-mail: jmadron@richmond.edu}\\ %
    \scriptsize University of Richmond, Digital Scholarship Lab %
\and Robert Nelson\thanks{e-mail: rnelson2@richmond.edu}\\ %
    \scriptsize University of Richmond, Digital Scholarship Lab %
\and Lauren Tilton\thanks{e-mail: ltilton@richmond.edu}\\ %
    \scriptsize University of Richmond, Rhetoric and Communications} %

\abstract{
  Museums, libraries, and other cultural institutions continue to prioritize
  and build web-based visualization systems that increase access and discovery to
  digitized archives. Prominent examples exist that illustrate impressive visualizations
  of a particular feature of a collection. For example, interactive maps showing
  geographic spread or timelines capturing the temporal aspects of collections.
  By way of a case study, this paper presents a new web-based visualization system that allows users to
  simultaneously explore a large collection of images along several different
  dimensions---spatial, temporal, visual, textual, and through additional metadata
  fields including the photographer name---guided by the concept of generous interfaces. The case study is a complete redesign of a previously released digital, public humanities project called Photogrammar (2014).
  The paper highlights the redesign's interactive visualizations that are now
  possible by the
  affordances of newly available software. All of the code is open-source in order to
  allow for re-use of the codebase to other collections with a similar structure.
}

\CCScatlist{
  \CCScatTwelve{Human-centered computing}{Visu\-al\-iza\-tion}{Information visualization}{};
  \CCScatTwelve{Human-centered computing}{Visu\-al\-iza\-tion}{Geographic visualization}{}
}

\begin{document}

\firstsection{Introduction}

\maketitle

Over the past few decades, cultural institutions have made extensive commitments to digitizing
and making their collections available over the web. While some collections are only
able to provide  limited access to metadata or through an institutional login screen for
legal reasons, many archives have been able to open up digitized materials to the public
at large. Prominent examples of institutions that have made large collections digitally
available include the Rijksmuseum, the British Library, the Library of Congress, the
Metropolitan Museum of Art, and the Smithsonian \cite{dijkshoorn2018rijksmuseum, alexiev2018museum}.
Making materials digitally available is an exciting step in improving access and discovery.
However, additional contextual framing and digital tools are often needed to guide users through
the material in order to engage with a broad public audience.

In response to the increase in access to digital collections, experts across various fields including digital humanities and data science have begun to think of collections as data \cite{padilla2018collections}.  Many tools focus on a specific mode of analysis in order to explore collections from a
specific point of view. For example, the Knight Lab's \textit{Timeline.js} provides a
library for building custom visualizations of data distributed over time. Tools created by
companies such as ArcGIS and Carto provide similar options for spatial data. Textual data
can be visualized through the use of keywords (e.g., \textit{Voyant}) or
topic models (e.g., \textit{jsLDA}).
General metadata fields can be visualized with libraries such as \textit{d3} and
\textit{cube.js}. While not
specifically designed for web-base usage, Gephi provides a range of options for showing
and exploring network relationships. The WebGL-based library
\textit{PixPlot} \cite{pixplot} provides an impressive framework for exploring large
photographic collections using state-of-the-art neural network models. In most of
these cases, data must be loaded into a series of tools to visualize the data
from different angles and for different purposes.

Significantly less attention has been paid to approaches for visualizing many unstructured
components of a collection at the same time. Digital storytelling applications, such as
the knight lab's \textit{StoryMap.js}, have shown the power of putting together images,
text, and other metadata within a specific narrative structure. However, web-based tools
for the open exploration of such collections are limited. Exploring the many dimensions of
multimodal collections generally requires the use of programming software, which has a
significantly higher barrier to entry that limits its audience. This hampers the ability of
cultural institutions to allow digital visitors of their collections to understand the
subtle interplay of an archive's components.


In this paper, we present a web-based
visualization system designed with the concept of generous interfaces for a collection of over 170,000 historic photographs through a case study. Photogrammar, a digital public humanities project and our case study, builds off of
our previous work with the collection to allow users to explore spatial, temporal, textual,
visual, and structured components of the collection within a single interactive generous interface.
This significantly increases the kinds of analysis available to users directly within the
site while visually communicating the interconnectedness of data and metadata.

The remainder of the article is organized as follows. We start with a brief history
of the archive we are working with and describing our original digital website (published
in 2014) along with its strengths and challenges. We then present the integrated design of
the new site and illustrate the new analyses afforded by the redesign. Particular attention
is given to the use of newly available image analysis algorithms and how they have changed
the possibilities for visualizing and therefore accessing and interpreting material.
A concluding section discusses
open challenges and our proposed path for developing a general-purpose visualization
library from our specific case study.

\begin{figure}[!ht]
 \centering
 \includegraphics[width=1.0\columnwidth]{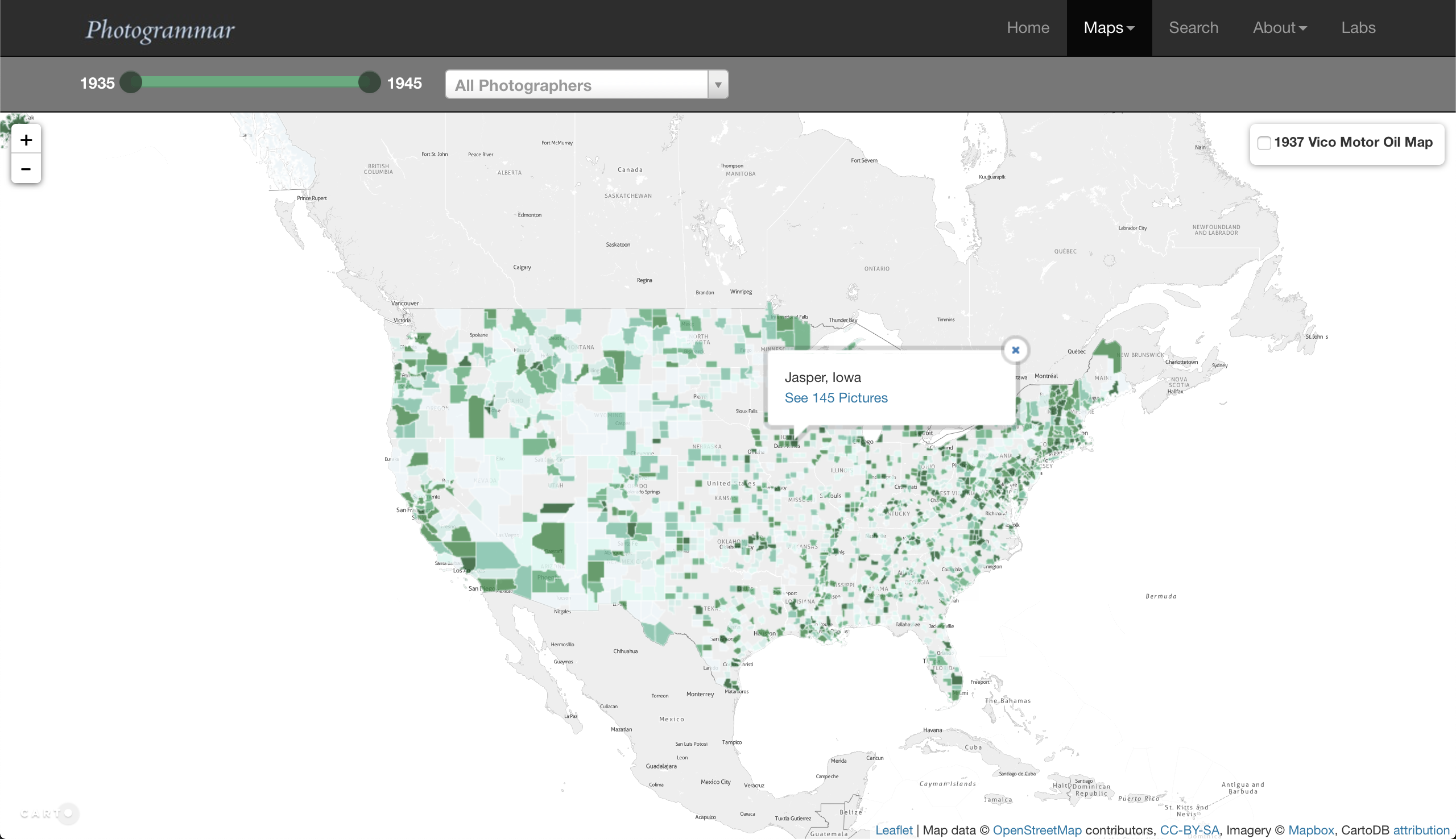}
 \caption{Original interactive map from the \textit{Photogrammar} website
 (2014 through mid-2020), implemented using slippy maptiles. County-level objects show
 the number of photographs
 taken in each region; the photographs can be subset and the counts updates by using the
 pull-down menu to select a photographer or the timeline to select a range of years.
 Clicking on a county---as shown here for Jasper, Iowa---reveals the exact number of
 photographs and a provides a link to a search page.
 }
 \label{fig:orig-map}
\end{figure}

\begin{figure}[!ht]
 \centering
 \includegraphics[width=1.0\columnwidth]{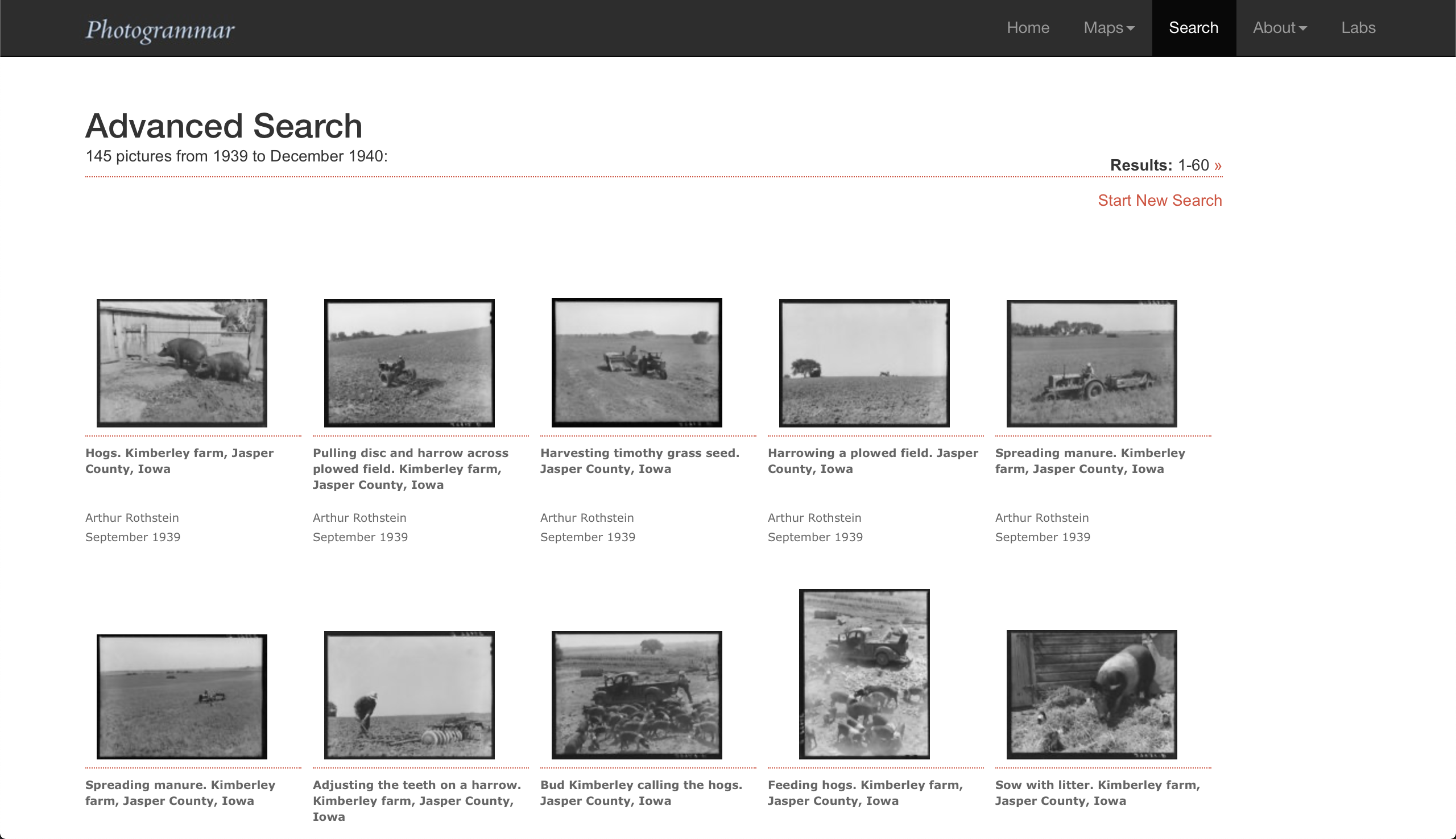}
 \caption{Search page on the original \textit{Photogrammar} website (2014 through mid-2020),
 showing the results of clicking on the link in Figure~\ref{fig:orig-map}. Thumbnails are
 displayed in a grid along with limited metadata (location, photographer, month, and year).
 Images are displayed in groups of 60. A more specific faceted search can be run by clicking
 on the `Start New Search' prompt in the upper-lefthand corner of the page.}
 \label{fig:orig-search}
\end{figure}

\begin{figure}[!ht]
 \centering
 \includegraphics[width=1.0\columnwidth]{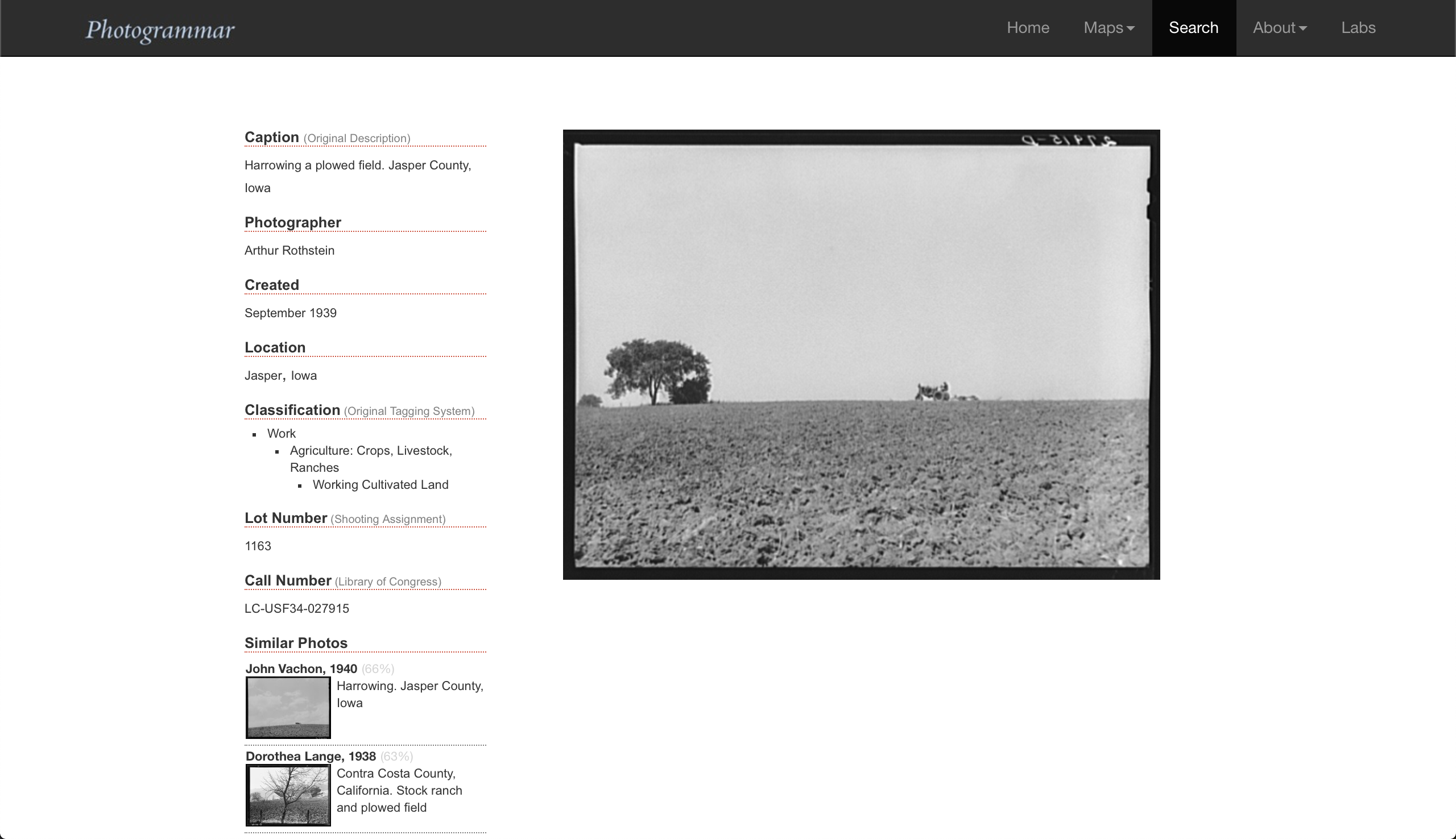}
 \caption{Photograph page from the original \textit{Photogrammar} website (2014 through
 mid-2020). This is the result of clicking on the fourth image shown in the results from
 Figure~\ref{fig:orig-search}. Most of the metadata fields are hyperlinks to further
 searches to show other images from the same location, topic, or by the same photographer.
 A set of similar photographs are suggested based on the similarity of the image captions.}
 \label{fig:orig-image}
\end{figure}

\section{Background: FSA-OWI and Photogrammar}

The FSA-OWI photographic archive is the result of a commission by the
United States Farm Security Administration and Office of War Information
(FSA-OWI) to document American life during the Great Depression and
World War II \cite{baldwin1968poverty}.
Physically housed in the Library of Congress (LoC),
the archive serves as an important visual record for scholars across
the academy and the public-at-large. The collection includes many iconic photographs,
include Dorothea Lange's \textit{Migrant Mother}, Gordon Parks' \textit{American Gothic},
and Arthur Rothstein's photos of the American Dust Bowl. The collection's
170,000 monochrome and color photographs resides entirely
within the public domain and offers a unique snapshot of the nation in the
1930s and 1940s \cite{trachtenberg1990reading}.

From 2012 to 2016, Arnold and Tilton, in collaboration with Laura Wexler, constructed an
interactive website called \textit{Photogrammar} as a new approach to accessing and
visualizing the FSA-OWI photographic collection. The full team came to include Trip Kirkpatrick, Peter Leonard, Stacey Maples, and Ken Panko.
The project began as a response to the challenges of navigating the digital
and physical archive at the LoC. Users either
had to search through the LoC's filing cabinets of over 88,000 prints or
LoC's online site, which allows for limited exploration. Through collaboration
with the collection's curator to harness the collection's rich metadata, the project
developed creative and innovative new approaches to the collection using digital and public
humanities methods. By cleaning and transforming the metadata, standardized the fields and
geotagged the collection, the site allows users to map photographs, run faceted searches,
and see historic cataloging systems used to organize the physical collection\cite{arnold2017dhq}.

The original incarnation of the \textit{Photogrammar} project can be seen in
Figures~\ref{fig:orig-map} through \ref{fig:orig-image}. The core of the site was
implemented as a MySQL database containing records for every photograph in the collection.
One page implemented an advanced faceted search, Figure~\ref{fig:orig-search}, which would
produce a sorted list of results on a new webpage. Clicking on a result opened a new
dynamically produced record page, Figure~\ref{fig:orig-image}, displaying a large version
of the image and associated metadata. This interface offered a similar search experience to
the existing site at the Library of Congress, while providing more control over the search
parameters and improved cleaning of the available metadata \cite{arnold2017}. Most users,
however, entered the collection through one of the available interactive visualizations
on the site. These include an interactive map, Figure~\ref{fig:orig-map}, a crosstab
metadata explorer, and a hierarchical description of an historic classification system
used by the Library of Congress. These visualization provided different aggregate views of
the 170,000 photographs. They are coded using individual JavaScript libraries; access to
the individual photographs was provided by pre-filled links to the advanced search
interface.

The original version of the \textit{Photogrammar} website provided many different ways of
viewing the collection, opening up the FSA-OWI collection to interactive exploration to a
wide public audience. In our redesign, described in the remainder of this article, our
goal was to build on new technology stacks to address several design issues in the original
site, with a particular focus on integrating the various search and exploration pages into
a single integrated web-based visualization of the archive.

\begin{figure*}[!ht]
 \centering
 \includegraphics[width=0.95\textwidth]{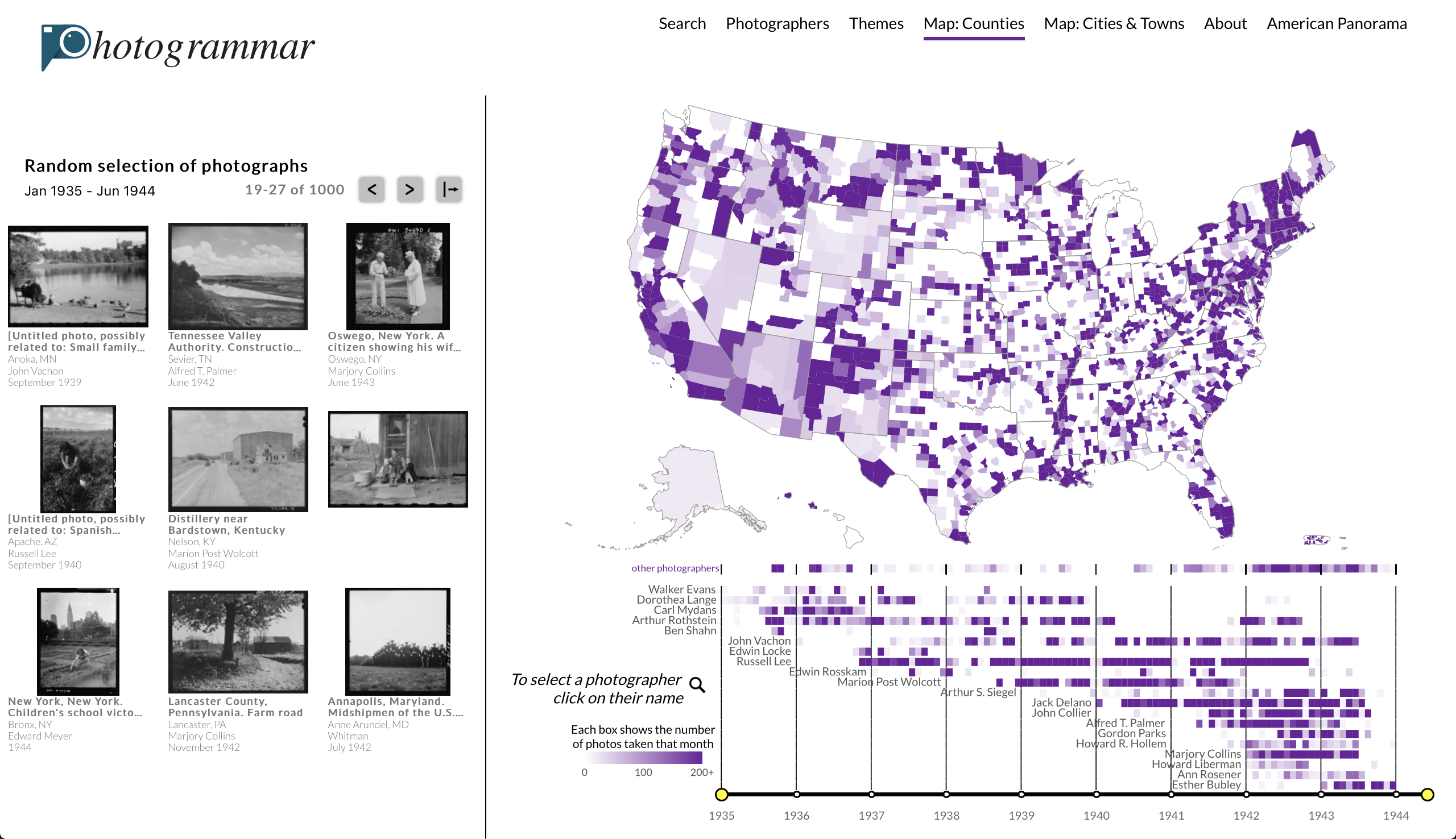}
 \caption{Landing page of the new \textit{Photogrammar} website, which shows the currently
 selected images (left), a chloropleth map (right), and selected photographers and years
 (bottom). The elements are linked together to allow users to explore the interactions of
 different dimensions of the collection simultaneously.}
 \label{fig:new-landing}
\end{figure*}

\begin{figure*}[!ht]
 \centering
 \includegraphics[width=0.95\textwidth]{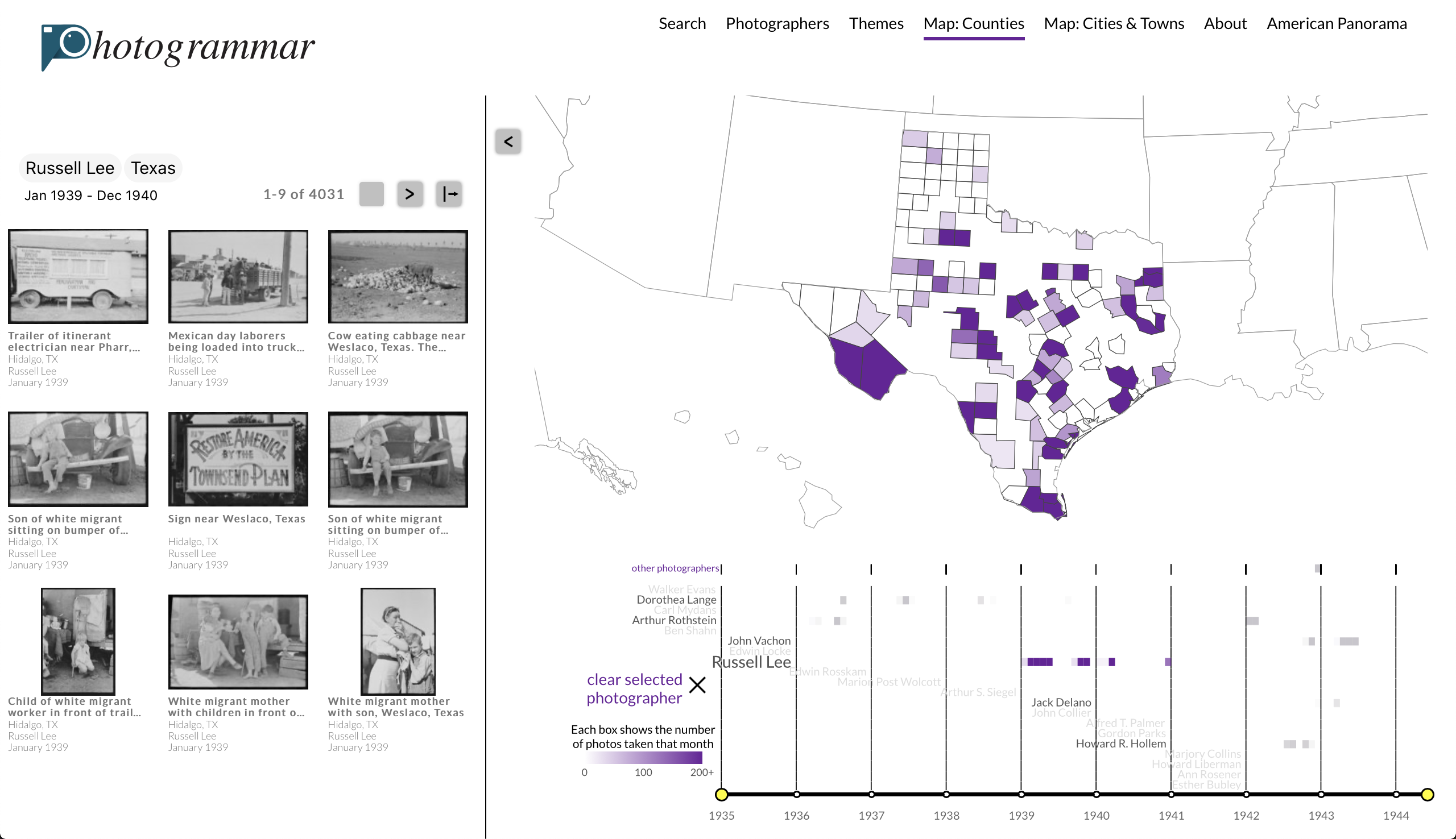}
 \caption{The result of clicking on the state of Texas and photographer Russell Lee on the
 landing page shown in Figure~\ref{fig:new-landing}. All three elements---the selected
 photographs, and spatial-temporal chloropleth visualizations---are updated to reflect only
 images by Russell Lee taken in Texas.}
 \label{fig:new-spatial-search}
\end{figure*}

\begin{figure}[!ht]
 \centering
 \includegraphics[width=1.0\columnwidth]{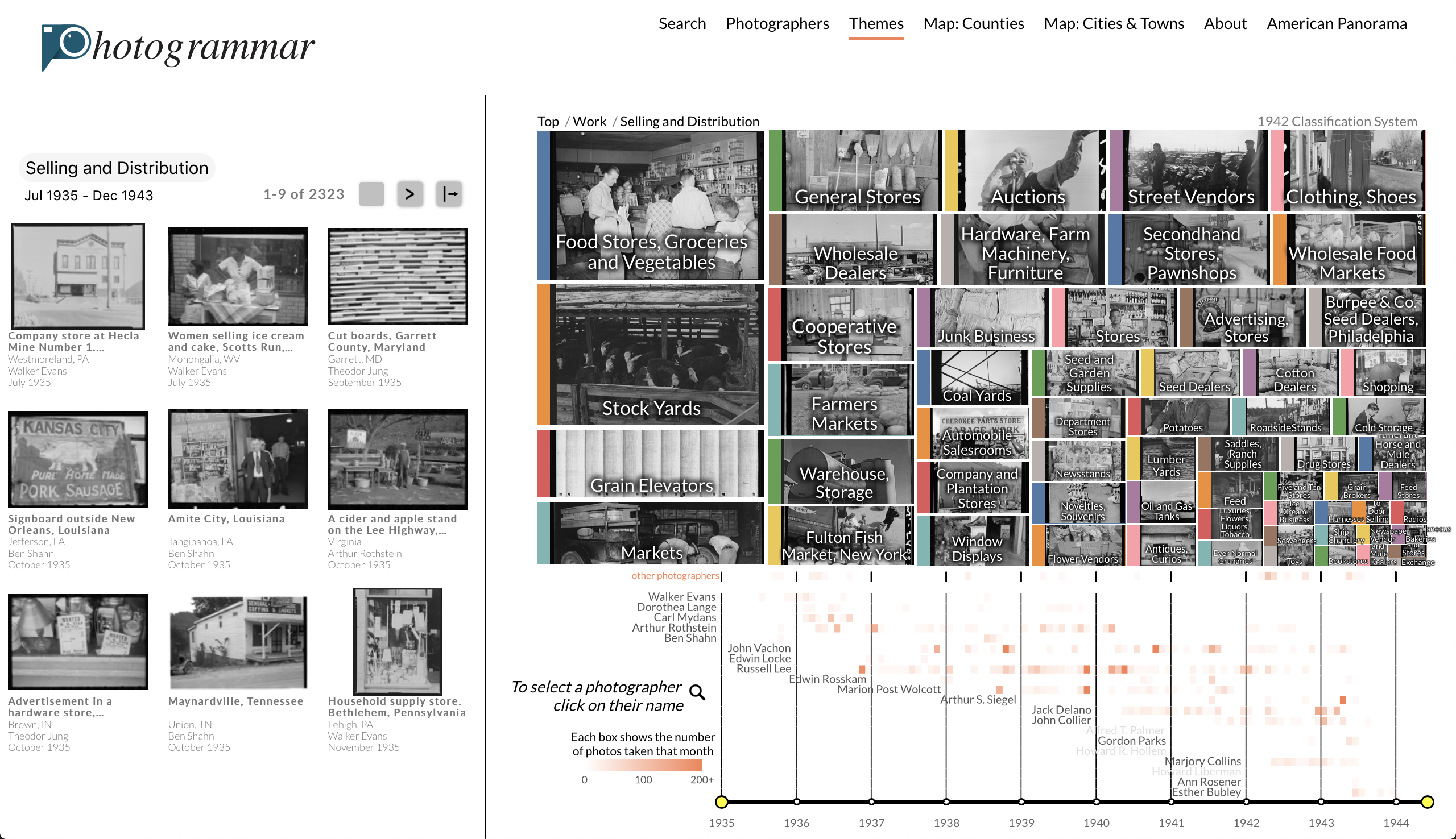}
 \caption{The map in Figure~\ref{fig:new-landing} can be replaced by other visualizations
 by selecting an option at the top. This figure shows a hierarchical treemap of the historic
 themes assigned to the images in the collection by Paul Vanderbilt. As with the map, the
 selected photos on the left and interactive timeline at the bottom interactively update
 with the selected Themes.}
 \label{fig:new-themes}
\end{figure}

\begin{figure}[!ht]
 \centering
 \includegraphics[width=1.0\columnwidth]{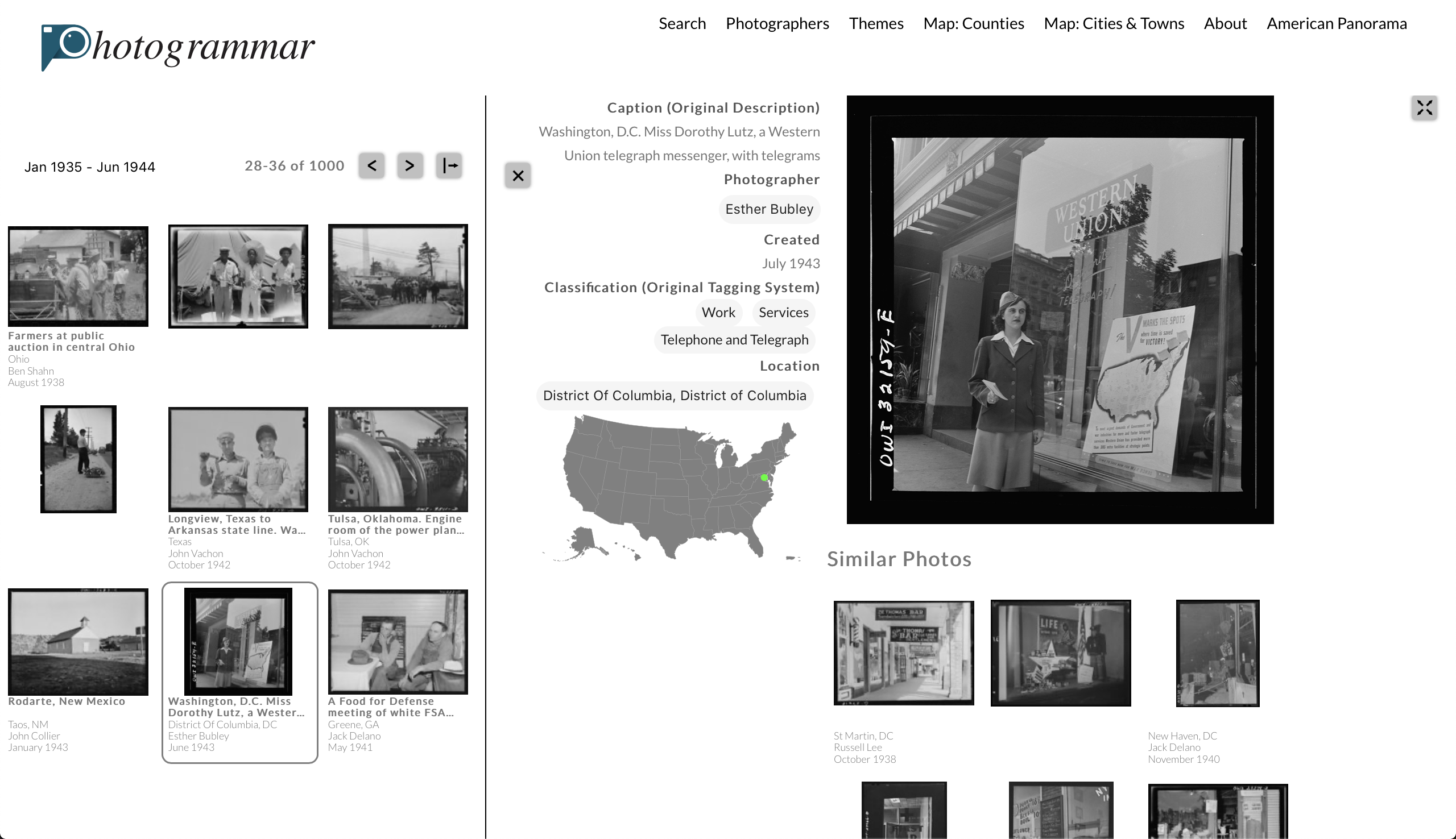}
 \caption{The result of clicking on an image in the new \textit{Photogrammar} website.
 A larger version of the image is shown, along with metadata and similar photographs. A
 small map allows users to retain the spatial dimension of the collection.}
 \label{fig:new-recs}
\end{figure}

\section{Generous Interfaces and the Digital Public Humanities}

The concept of a `generous interface' was introduced by Mitchell Whitelaw in 2015
\cite{whitelaw2017}. Focused on the
collections held by cultural heritage institutions, he called for a conceptual shift from
the singular, narrow interaction of the keyword search (usually in the form of a blank box such as Google Search)  to interfaces that communicate the scale, scope, and richness of a
collection. Visualizations are a key component and a strategy. Centering exploration,
interactivity, and scalability, generous interfaces are an ethos, praxis, and outcome.

Generous interfaces can also be understood as a strategy for the digital, public humanities (DPH).
The rapidly expanding set of projects online that identify as DPH---projects that seek to
engage, and often co-create, with a set of publics beyond the academy and commonly are built
to expand access and interpretation using methods from the digital humanities---has brought
questions about how to engage with audiences and users to the fore. As Sheila Brennan
argues, ``It is important to recognize that projects and research may be available online,
but that status does not inherently make the work digital public humanities or public
digital humanities" \cite{brennan2016}.
As Jordana Cox and Lauren Tilton have argued, DPH can ``forefront
visual, nonlinear, and interactive argumentation in oder to engage publics in generative
humanistic inquiry" \cite{tilton2019}.  One way is  projects designed with generous
interfaces that further generative and generous humanistic inquiry.

Through a more detailed explanation of the redesign of Photogrammar, we demonstrate how
generous interfaces and DPH are theoretical frameworks and methods for accessing and
interpreting cultural heritage data through interactive visualizations. The new version of Photogrammar forges an integrated approach to generous interfaces.
Rather than treat each interactive visualization as a
discrete task, we demonstrate how we can connect different types of visualizations with
concepts from information retrieval such as search, browse, and recommender systems. What
is considered data and metadata shifts based on the community and their interests;
therefore, we show how toggling between and integrating `data' and `metadata' offers a
strategy for creating generous interfaces built around integrated visualizations that
further access and interpretation of collections.

\section{A More Generous Redesign}

The new version of \textit{Photogrammar} provides a complete redesign of the web-based
visualization system, which offers a more integrated user experience through a generous
interface built on a modern technology stack. Figure~\ref{fig:new-landing} shows the new
landing page for the site, featuring a selection of photographic thumbnails (left), an
interactive map (top right), and interactive timeline split by photographer (bottom right).
The entire page is now implemented as a \textit{React.js} app, allowing all of the elements of the
page to be linked together. Clicking on the state of Texas on the map, for example, zooms
to the state of Texas. At the same time, the photographs on the left and timeline at the
bottom are updated simultaneously. Further clicking on a photographer at the bottom updates
the map and photos further. The result of clicking on the state of Texas and photographer
Russell Lee is given in Figure~\ref{fig:new-spatial-search}. Linking these elements
together helps to center the photographs  themselves, the main attraction for most users,
by putting them on the landing page and updating them dynamically rather than through a
secondary page. It also allows for transdisclinary analyses at the intersection of space,
time, photographer, and themes that were not available through the original site.

A major limiting factor in offering many coupled visualizations is the available screen
area available for multiple visualizations. Significant effort has been put into compressing
the amount of space needed for the default visualizations. For example, the dynamic slippy
maptile approach of the original site has been replaced with a fixed zoom level map served
by a GeoJSON file \cite{horbinski2020use}. To compress space and improve the visibility of
the entire collection, Hawaii, Alaska, Puerto Rico, and the U.S. Virgin Islands have been
repositioned into the main map. For the timeline of photographers, only the most prolific
photographers are included in the default visualization. Additional photographers may
still be selected, however, by clicking the `other photographers' button.

Even with these space saving efforts, it is not possible to include all of the
visualizations in a single window that is still visible on most standard computer monitors.
In order to access other visualizations, the default interactive map can be replaced with
a different visualization by selecting one of the options at the top of the page. These
include a more granular points-based map and a theme visualization
(Figure~\ref{fig:new-themes}). As with the original map, all of these visualizations are
coupled with the timeline and search results on the side bar.

The search results interactively within the page as the maps, timelines, and themes are
selected provide a way of making visible the images themselves from the beginning. The
search results on the side bar contain some minimal metadata, and a small thumbnail of the
image. Pages of the results can be scrolled through to see all of the relevant images.
It is still needed, however, to use more screen space to see a larger version of the image
and all of the associated metadata. In order to not lose the search results, and to visually
emphasis that all of the elements (i.e. metadata and photos) are connected, this graphic is overlaid over the map and
timeline on the right-hand side of the page. An example is shown in
Figure~\ref{fig:new-recs}. In order to continue to stress the connection to the spatial
visualization, a small in-lay map of the United States is shown within the photograph page. By aggregating and connecting different type of visualizations, the system creates a generous interface that amplifies digital public humanities inquiry into a collection.

\section{Recommendation Systems}

The interactive maps, timeline, and theme treeplot are all designed to allow users of the
website to explicitly select subsets of the FSA-OWI collection that they are interested in
viewing. Typical behavior involves selection a region of the map that a user is most
familiar with---such as their current location or a hometown---or selecting one of the
better known photographers. Opening familiar lines of search for users has many benefits:
it creates a personal connection between the (meta)data and the user, it allows users to bring
their own knowledge into their interpretation of the collection, and it increases the
level of user engagement. The downside of this way into the archive, however, is that it
risks getting users stuck in a small subset of the larger collection. In order to overcome
this limitation, \textit{Photogrammar} also incorporates a set of recommendations that
cut across the fixed metadata fields that drives the initial discovery step of exploring
each collection.

A recommendation system is an algorithm for predicting what records in a collection a user
may be interested in \cite{park2012literature}. These systems feature prominently in
commercial services. For example,
product pages on Amazon features other `similar products', Netflix recommends new releases
of interest, and Spotify creates artifically constructed playlists based on prior listening
behavior. In the context of a digital archive, recommendation systems are most commonly used
to recommend records that are most similar to the particular record that a user is currently
looking at. It is possible to determine the most similar records based on the search
behavior of other users. This method for building a recommendation system, a process
called \textit{collaborative} filtering, requires a large amount of data, creates
privacy concerns, and necessitates constantly updating models based on changing behaviors
\cite{calandrino2011you}.
An alternative method, \textit{content} filtering, bases recommendations entirely on the
content of the records themselves \cite{thorat2015survey}. We utilize content filtering in
our generous interface of the FSA-OWI archive.

The initial version of the \textit{Photogrammar} website utilized a recommendation system
that recommended similar records on the individual photograph pages. The interface is shown
at the bottom of Figure~\ref{fig:orig-image}. These similarities are based on using the
text in the captions of the photographs. Specifically, we compute term
frequency-inverse document frequency (TF-IDF) scores and identify captions that are the most
similar to the caption of interest. Pervious work has discussed the various iterations that
this system has gone through. For example, a process of named entity recognition (NER) was
used to remove geographic names from the captions \cite{arnold2017}. A word-embedding model
was used to to find similar captions that use slightly different words, such as `apple
orchard' and `pear trees' \cite{arnold2018}. This caption-based system helps find
connections that would otherwise be hidden through other interfaces. However, it does have
some limitations. Less than half of the images have captions, and many of these are too
short to be useful for recommendation purposes. Furthermore, the captions only describe
one main object of study and rarely
mention elements of style and composition. In order to capture these elements, a different
approach is needed.

The new \textit{Photogrammar} website implements a content filtering recommendation
system based on visual similarity. Every image in the collection was projected into the
penultimate layer of a large convolutional neural network in order to obtain an embedding
of the entire collection into a high-dimensional space \cite{szegedy2015going}. Each record
is then associated with other photographs that have the most similar set of embedding
values \cite{arnold2019}. This method
tends to create connections between images that contain similar objects, events, or share
similar forms of composition \cite{pixplot}. While the neural network used is trained
entirely on color images, the implicit connections created by the embedding produce
surprisingly consistent and meaningful results on the predominantly black-and-white images
in the FSA-OWI collection \cite{russakovsky2015imagenet}. The new recommendation system is
shown in the bottom of
Figure~\ref{fig:new-recs}. The visual search method overcomes many of the difficulties of
the entirely text-based method, such as being able to produce recommendations for all
images in the corpus and finding links between elements of the images not explicitly
mentioned in each caption. The new recommender system also responds to consistent feedback from users who asked about visual search.

Computing embedding-based visual similarities for each image is a computationally intensive
task. Rather than running this each time a user requests a particular image, the graph of
similar images has instead been precomputed and stored as static files on the website. In
addition to the benefit of reducing the computational load of the site, this approach also
provides a simple path forward for further iterations of the recommendation system, such as
integrating joint text-image embeddings
\cite{alikhani-etal-2019-cite, singhal-etal-2019-learning}. A new recommendation system can
be pushed into the site by simply updating the static files describing the most similar
photographs.

\section{Conclusions and Future Work}

The new, completely redesigned version of the \textit{Photogrammar} web-based visualization
and search interface offers numerous improvements over the already successful original
version of the site. In line with the goals of the digital public humanities and generous
design, the new site allows for a more integrated user-experience with a number of tightly
connected elements that all interactively update with one another. Additional modifications
built into the new maps, timeline, and recommendation system further improve the ability of
users to explore and connect with the photographs from the FSA-OWI archive. As a result, the visualizations come together, harnessing the relationships between and across the metadata, to realize a generous interface for the photographs. At the same time, the visualizations allows the user to toggle between and across different objects of study. Metadata fields, as classified by the Library of Congress and through the structure of the database serving Photogrammar, can become the object of study, such as interest in the work of a specific photographer such as Marion Post Wolcott or a state such as as Texas, and therefore the `data'. Generous interfaces, therefore, provide a theoretical and methodological frame for designing systems that disrupt classifications of data and metadata in the way that humanistic inquiry often demands.

In our redesign of the site, a key concern was building maintainable and reusable
components. Future work on the project will focus on decoupling the elements specific to
this collection (such as the theme visualizer) and turning the codebase into a documented
framework that can be used by other archival projects in the digital public humanities
that feature rich, multimodal data.

\nocite{*}

\bibliographystyle{abbrv-doi}

\bibliography{photogrammar-viz-paper}

\begin{thebibliography}{10}

\bibitem{abdollahpouri2019managing}
H.~Abdollahpouri, R.~Burke, and B.~Mobasher.
\newblock Managing popularity bias in recommender systems with personalized
  re-ranking.
\newblock In {\em The Thirty-Second International Flairs Conference}, 2019.

\bibitem{alexiev2018museum}
V.~Alexiev.
\newblock Museum linked open data: Ontologies, datasets, projects.
\newblock {\em Digital Presentation and Preservation of Cultural and Scientific
  Heritage}, (VIII):19--50, 2018.

\bibitem{alikhani-etal-2019-cite}
M.~Alikhani, S.~Nag~Chowdhury, G.~de~Melo, and M.~Stone.
\newblock {CITE}: A corpus of image-text discourse relations.
\newblock In {\em Proceedings of the 2019 Conference of the North {A}merican
  Chapter of the Association for Computational Linguistics: Human Language
  Technologies, Volume 1 (Long and Short Papers)}, pp. 570--575. Association
  for Computational Linguistics, Minneapolis, Minnesota, June 2019. doi: {{%
10\hspace{.1pt}\discretionary{.}{%
}{.}\hspace{.4pt}18653\discretionary{/}{%
}{/}v1\discretionary{/}{%
}{/}N19\discretionary{%
}{-}{-}1056}}


\bibitem{arnold2017}
T.~Arnold, P.~Leonard, and L.~Tilton.
\newblock Knowledge creation through recommender systems.
\newblock {\em Digital Scholarship in the Humanities}, 32(3):151--157, 2017.

\bibitem{arnold2017dhq}
T.~Arnold, S.~Maples, L.~Tilton, and L.~Wexler.
\newblock Uncovering latent metadata in the {FSA-OWI} photographic archive.
\newblock {\em {DHQ}: Digital Humanities Quarterly}, 11(2), 2017.

\bibitem{arnold2018}
T.~Arnold and L.~Tilton.
\newblock Cross-discourse and multilingual exploration of textual corpora with
  the {DualNeighbors} algorithm.
\newblock pp. 50--59, 2018.

\bibitem{arnold2019}
T.~Arnold and L.~Tilton.
\newblock Distant viewing: Analyzing large visual corpora.
\newblock {\em Digital Scholarship in the Humanities}, 34(1):i3--i16, 2019.

\bibitem{baldwin1968poverty}
S.~Baldwin.
\newblock Poverty and politics; the rise and decline of the farm security
  administration.
\newblock 1968.

\bibitem{brennan2016}
S.~A. Brennan.
\newblock Public, first.
\newblock In M.~Gold and L.~Klein, eds., {\em Debates in the Digital
  Humanities}. University of Minnesota Press, 2016.

\bibitem{buolamwini2018gender}
J.~Buolamwini and T.~Gebru.
\newblock Gender shades: Intersectional accuracy disparities in commercial
  gender classification.
\newblock In {\em Conference on fairness, accountability and transparency}, pp.
  77--91, 2018.

\bibitem{caesar2018coco}
H.~Caesar, J.~Uijlings, and V.~Ferrari.
\newblock {COCO}-stuff: Thing and stuff classes in context.
\newblock In {\em Proceedings of the IEEE Conference on Computer Vision and
  Pattern Recognition}, pp. 1209--1218, 2018.

\bibitem{calandrino2011you}
J.~A. Calandrino, A.~Kilzer, A.~Narayanan, E.~W. Felten, and V.~Shmatikov.
\newblock You might also like: Privacy risks of collaborative filtering.
\newblock In {\em 2011 IEEE symposium on security and privacy}, pp. 231--246.
  IEEE, 2011.

\bibitem{concordia2009not}
C.~Concordia, S.~Gradmann, and S.~Siebinga.
\newblock Not (just) a repository, nor (just) a digital library, nor (just) a
  portal: A portrait of {E}uropeana as an {API}.
\newblock In {\em World Library and Information Congress: 75th IFLA General
  Conference and Council}, 2009.

\bibitem{tilton2019}
J.~Cox and L.~Tilton.
\newblock The digital public humanities: Giving new arguments and new ways to
  argue.
\newblock vol.~19, pp. 127--146. 2016.

\bibitem{dijkshoorn2018rijksmuseum}
C.~Dijkshoorn, L.~Jongma, L.~Aroyo, J.~Van~Ossenbruggen, G.~Schreiber,
  W.~Ter~Weele, and J.~Wielemaker.
\newblock The {R}ijksmuseum collection as linked data.
\newblock {\em Semantic Web}, 9(2):221--230, 2018.

\bibitem{pixplot}
D.~Duhaime.
\newblock {PixPlot}: Visualize large image collections with {WebGL}.
\newblock \url{https://github.com/YaleDHLab/pix-plot}, 2019.

\bibitem{guha2016schema}
R.~V. Guha, D.~Brickley, and S.~Macbeth.
\newblock {Schema.org}: evolution of structured data on the web.
\newblock {\em Communications of the ACM}, 59(2):44--51, 2016.

\bibitem{horbinski2020use}
T.~Horbi{\'n}ski and D.~Lorek.
\newblock The use of {Leaflet} and {GeoJSON} files for creating the interactive
  web map of the preindustrial state of the natural environment.
\newblock {\em Journal of Spatial Science}, pp. 1--17, 2020.

\bibitem{khaw2019individual}
M.~W. Khaw, L.~Stevens, and M.~Woodford.
\newblock Individual differences in the perception of probability.
\newblock {\em Available at SSRN 3446790}, 2019.

\bibitem{kirillov2019panoptic}
A.~Kirillov, K.~He, R.~Girshick, C.~Rother, and P.~Doll{\'a}r.
\newblock Panoptic segmentation.
\newblock In {\em Proceedings of the IEEE conference on computer vision and
  pattern recognition}, pp. 9404--9413, 2019.

\bibitem{kumar2016comparative}
A.~Kumar and R.~K. Singh.
\newblock Comparative analysis of {Angular.js} and {React.js}.
\newblock {\em International Journal of Latest Trends in Engineering and
  Technology}, 7(4):225--227, 2016.

\bibitem{lin2015bilinear}
T.-Y. Lin, A.~RoyChowdhury, and S.~Maji.
\newblock Bilinear {CNN} models for fine-grained visual recognition.
\newblock In {\em Proceedings of the IEEE international conference on computer
  vision}, pp. 1449--1457, 2015.

\bibitem{mcauley2015image}
J.~McAuley, C.~Targett, Q.~Shi, and A.~Van Den~Hengel.
\newblock Image-based recommendations on styles and substitutes.
\newblock In {\em Proceedings of the 38th International ACM SIGIR Conference on
  Research and Development in Information Retrieval}, pp. 43--52, 2015.

\bibitem{mensink2014rijksmuseum}
T.~Mensink and J.~Van~Gemert.
\newblock The {R}ijksmuseum challenge: Museum-centered visual recognition.
\newblock In {\em Proceedings of International Conference on Multimedia
  Retrieval}, pp. 451--454, 2014.

\bibitem{miyakita2019exploring}
G.~Miyakita, S.~Arima, M.~Yasui, and K.~Okawa.
\newblock Exploring digital cultural heritage beyond moocs: Design, use, and
  efficiency of generous interfaces.
\newblock In {\em 2019 IEEE Learning With MOOCS, LWMOOCS 2019}, pp. 42--46.
  Institute of Electrical and Electronics Engineers Inc., 2019.

\bibitem{padilla2018collections}
T.~G. Padilla.
\newblock Collections as data: Implications for enclosure.
\newblock {\em College and Research Libraries News}, 79(6):296, 2018.

\bibitem{park2012literature}
D.~H. Park, H.~K. Kim, I.~Y. Choi, and J.~K. Kim.
\newblock A literature review and classification of recommender systems
  research.
\newblock {\em Expert systems with applications}, 39(11):10059--10072, 2012.

\bibitem{platt1999using}
J.~C. Platt.
\newblock Using analytic {QP} and sparseness to speed training of support
  vector machines.
\newblock In {\em Advances in neural information processing systems}, pp.
  557--563, 1999.

\bibitem{russakovsky2015imagenet}
O.~Russakovsky, J.~Deng, H.~Su, J.~Krause, S.~Satheesh, S.~Ma, Z.~Huang,
  A.~Karpathy, A.~Khosla, M.~Bernstein, et~al.
\newblock {ImageNet} large scale visual recognition challenge.
\newblock {\em International journal of computer vision}, 115(3):211--252,
  2015.

\bibitem{seitsonen2017crowdsourcing}
O.~Seitsonen.
\newblock Crowdsourcing cultural heritage: public participation and conflict
  legacy in {F}inland.
\newblock {\em Journal of Community Archaeology \& Heritage}, 4(2):115--130,
  2017.

\bibitem{singhal-etal-2019-learning}
K.~Singhal, K.~Raman, and B.~ten Cate.
\newblock Learning multilingual word embeddings using image-text data.
\newblock In {\em Proceedings of the Second Workshop on Shortcomings in Vision
  and Language}, pp. 68--77. Association for Computational Linguistics,
  Minneapolis, Minnesota, June 2019. doi: {{%
10\hspace{.1pt}\discretionary{.}{%
}{.}\hspace{.4pt}18653\discretionary{/}{%
}{/}v1\discretionary{/}{%
}{/}W19\discretionary{%
}{-}{-}1807}}


\bibitem{speakman2018user}
R.~Speakman, M.~M. Hall, and D.~Walsh.
\newblock User engagement with generous interfaces for digital cultural
  heritage.
\newblock In {\em International Conference on Theory and Practice of Digital
  Libraries}, pp. 186--191. Springer, 2018.

\bibitem{szegedy2015going}
C.~Szegedy, W.~Liu, Y.~Jia, P.~Sermanet, S.~Reed, D.~Anguelov, D.~Erhan,
  V.~Vanhoucke, and A.~Rabinovich.
\newblock Going deeper with convolutions.
\newblock In {\em Proceedings of the IEEE conference on computer vision and
  pattern recognition}, pp. 1--9, 2015.

\bibitem{tagg2009disciplinary}
J.~Tagg.
\newblock {\em The disciplinary frame: Photographic truths and the capture of
  meaning}.
\newblock U of Minnesota Press, 2009.

\bibitem{thompson2017computational}
L.~Thompson and D.~Mimno.
\newblock Computational cut-ups: The influence of {D}ada.
\newblock {\em The Journal of Modern Periodical Studies}, 8(2):179--195, 2017.

\bibitem{thorat2015survey}
P.~B. Thorat, R.~Goudar, and S.~Barve.
\newblock Survey on collaborative filtering, content-based filtering and hybrid
  recommendation system.
\newblock {\em International Journal of Computer Applications}, 110(4):31--36,
  2015.

\bibitem{trachtenberg1990reading}
A.~Trachtenberg.
\newblock {\em Reading American Photographs: Images as History-{M}athew {B}rady
  to {W}alker {E}vans}.
\newblock Macmillan, London, England, 1990.

\bibitem{vinyals2016show}
O.~Vinyals, A.~Toshev, S.~Bengio, and D.~Erhan.
\newblock Show and tell: Lessons learned from the 2015 {MS COCO} image
  captioning challenge.
\newblock {\em IEEE transactions on pattern analysis and machine intelligence},
  39(4):652--663, 2016.

\bibitem{wevers2019visual}
M.~Wevers and T.~Smits.
\newblock The visual digital turn: Using neural networks to study historical
  images.
\newblock {\em Digital Scholarship in the Humanities}, 2019.

\bibitem{whitelaw2017}
M.~Whitelaw.
\newblock Generous interfaces for digital cultural collections.
\newblock {\em Digital Humanities Quarterly}, 9(1), 2015.

\bibitem{wu2019detectron2}
Y.~Wu, A.~Kirillov, F.~Massa, W.-Y. Lo, and R.~Girshick.
\newblock Detectron2.
\newblock \url{https://github.com/facebookresearch/detectron2}, 2019.

\bibitem{zimmer2015twitter}
M.~Zimmer.
\newblock The {T}witter archive at the library of congress: Challenges for
  information practice and information policy.
\newblock {\em First Monday}, 20(7), 2015.

\end{thebibliography}
\end{document}